\documentclass{ws-ijmpa}
\begin{document}
\markboth{S. K. Blau, M. Visser \& A. Wipf}
{Analytic results for the effective action}
\catchline{}{}{}
\title{ANALYTIC RESULTS FOR THE EFFECTIVE ACTION}
\author{ STEVEN K. BLAU\footnote{Present address [2009]: 
AIP, Washington DC; e-mail: {\sf sblau@aip.org}.} \,\,
and \, MATT VISSER\footnote{Present address [2009]: 
School of Mathematics, Statistics, and Operations Research, \hfil \newline
Victoria University of Wellington, New Zealand; e-mail: {\sf matt.visser@msor.vuw.ac.nz}; \hfil\newline
url: {\sf http://homepages.msor.vuw.ac.nz/$\sim$visser/   } 
}}

\address{
Theoretical Division,
Los Alamos National Laboratory,
Los Alamos, New Mexico 87545}

\author{
ANDREAS WIPF\footnote{Present address [2009]: 
Theoretisch-Physikalisches-Institut,
Friedrich-Schiller-Universit\"at Jena, e-mail:
07743 Jena, Germany,  e-mail: {\sf wipf@tpi.uni-jena.de};
url: {\sf  http://www2.uni-jena.de/$\sim$p5anwi/ }
 }}

\address{
Max Planck Institut f\"{u}r Physik und Astrophysik,
Werner Heisenberg Institut f\"{u}r Physik,
\\
D--8000 M\"{u}nchen 40, Germany}
\maketitle 
\pub{Received 12 July 1990}{}

\begin{abstract}
  Motivated by the seminal work of Schwinger, we obtain explicit
  closed form expressions for the one--loop effective action in a
  constant electromagnetic field.  We discuss both massive and
  massless charged scalars and spinors in two, three, and four
  dimensions.  Both strong field and weak field limits are calculable.
  The latter limit results in an asymptotic expansion whose first term
  reproduces the Euler--Heisenberg effective Lagrangian.  We use the
  zeta function renormalization prescription, and indicate its
  relationship to Schwinger's renormalized effective action.

\bigskip
\centerline{Published version: IJMPA {\bf 6} (1991) 5409--5433. }

\bigskip
\centerline{doi:10.1142/S0217751X91002549 }

\bigskip
\centerline{12 July 1990; arXiv-ed 16 July 2009; \LaTeX-ed \today}

\end{abstract}
\font\Bigfont=cmmi10 scaled\magstep3      
\def\z{\hbox{$\textfont1=\Bigfont \zeta$}}
\def\no{\noindent}
\def\t{\textstyle}
\def\s{\scriptstyle}
\def\half{{\textstyle {1\over2}}}
\def\threehalves{{\textstyle{3\over2}}}
\def\quarter{{\textstyle {1\over4}}}
\def\sixth{{\textstyle {1\over6}}}
\def\twelfth{{\textstyle {1\over12}}}
\def\twentyfourth{{\textstyle {1\over24}}}
\def\dslash#1{{#1\!\!\!\slash\,}}
\def\Seff{S_{\rm eff}}
\def\Leff{{\script L}_{\rm eff}}
\def\Eeff{E_{\rm eff}}
\def\cosech{\hbox{\rm cosech}}
\def\coth{\hbox{\rm coth}}
\def\Re{\hbox{\rm Re}}
\def\Im{\hbox{\rm Im}}
\def\integer{\hbox{\rm integer}}
\def\frac{\hbox{\rm frac}}
\def\infinity{\infty}
\def\curl{{\partial}}
\def\g5{{\gamma_5}}
\def\eps{\epsilon}
\def\etal{{\sl et al.}}
\def\eg{{\sl e.g.}}
\def\ie{{\sl i.e.}}
\def\Break{\hfil\break}
\def\dirac{{D\!\!\!\!\slash\,}}
\def\dsq{\dirac^2}
\def\Sum{\sum}
\def\script{\mathcal}
\def\tr{{\mathrm{tr}}}
\def\be{\begin{equation}}
\def\ee{\end{equation}}
\def\bea{\begin{eqnarray}}
\def\eea{\end{eqnarray}}

\section{Introduction}
 
In the path integral formulation of fermionic field theories, one is
forced to confront the determinant of the Dirac operator, while for
spin--0 bosons one encounters the determinant of the gauged Laplacian.
In the absence of gauge fields, and if the spacetime geometry is not
an issue, this determinant is an irrelevant constant.  However, there
is still a good deal of work to be done toward elucidating the
dependence of these determinants on background gauge fields and
gravitational fields.  These determinants are related to the one-loop
effective action via $\Seff\propto\ln\det D$, and, in the guise of the
one-loop effective action, have been the subject of considerable
efforts dating back at least to the seminal works of Euler and
Heisenberg,~\cite{Heisenberg} Weisskopf,~\cite{Weisskopf} and
Schwinger.~\cite{SchwingerGauge}
 
Only in two dimensions is the situation reasonably well understood.
The particularly simple geometry of compact two--dimensional Riemann
surfaces makes it possible to express the gauge--field--free Dirac
determinant in terms of Riemann theta functions and generalised
Dedekind eta
functions.~\cite{AlvarezGaume,AlvarezGaumeII,HokerII,HokerIII} On the
other hand, in the plane the determinant for localized gauge fields is
given by Schwinger's~\cite{Schwinger} famous result $\Seff\propto \int
A^\mu A_\mu$.  Combining the effects of non--trivial topologies and
non--zero field strength has been discussed in reference~\cite{bvw}.

In higher dimensions, only particular background fields can be
handled.  For example, the effect of conformal metric deformations on
the effective action is discussed in references~\cite{Birrell}
and~\cite{bvwII}. \,  Schwinger~\cite{SchwingerGauge} has considered the
effective action for constant electromagnetic field strength and for a
plane wave of electromagnetic radiation, both in flat
four--dimensional Minkowski space.  Four--dimensional instanton
solutions have also been considered.~\cite{Brown,Corrigan,Osborn}
 
In this paper we shall be interested in obtaining explicit analytic
results.  Accordingly, we are forced to restrict our attention to
particular backgrounds: flat spaces with zero--field or constant
field.  We use the zeta--function regularisation of determinants of
second-order elliptic operators.  The Dirac operator, $\dirac$, is
first order, but we shall define $\det \dirac \equiv
\sqrt{\det\dirac^2 }$.  In order that $\dirac^2$ may be an ellicptic
operator, we shall work in Wick rotated Euclidean spacetime.  In
section 2 we consider \hbox{(non--)Abelian} gauge fields defined on
multidimensional tori $(S^1)^d$.  The non--trivial topology of these
tori allows for the possibility of harmonic gauge potentials, that is,
potentials which have vanishing field strength but which are not pure
gauge.  The existence of these harmonic gauge potentials is associated
with the possibility of encountering non--trivial Wilson loops.  On
multidimensional tori, such gauge potentials are constant (up to a
gauge transformation), thus allowing explicit construction of the
eigenspectrum, zeta function, and effective action.  The dependence of
the effective action on these nontrivial Wilson loops may be viewed as
a generalized Aharonov--Bohm effect. In sections 3 through 7, we
consider gauge fields with constant field strength in arbitrarily many
dimensions, working our way up from two dimensions.  In all these
cases we shall determine the eigenspectra of the gauged Laplacian and
Dirac operator and shall calculate the associated zeta functions
explicitly.  Using special properties of zeta functions may often give
the determinant and effective Lagrangian density in closed form.
 
We shall also discuss the ``folk theorem'' $\Leff\sim B^{d/2} \ln B$,
and will point out a number of situations in which it is violated.
The physical implications of the logarithmic term are discussed in
reference~\cite{Notes}.  We shall show that this folk theorem is
generally true in even numbers of dimensions, though there are
exceptions, such as the scalar particle in two dimensions.  In odd
numbers of dimensions however, the logarithmic term is absent, and
generically we obtain $\Leff\sim B^{d/2}$.

\section{Field--Free Gauge Potentials}
 
In this section we shall see that a generalized Aharonov--Bohm effect
can influence the one-loop effective action of a system, even though
the electric field strength is everywhere zero.  Notice that an
Abelian gauge potential is gauge--equivalent to the sum of its
harmonic, monopole and coexact pieces
\begin{equation}
A=H+A^M+\delta\chi.
\label{E:decomp-betti}
\end{equation}
The number of independent harmonics equals the first Betti number and is
thus intimately related to the topology of the spacetime manifold.
We shall be interested in gauge potentials on $d$--dimensional
tori, in which case there are $d$ independent harmonics which may be
chosen to be constant one--forms. In this section we shall further
specialise to the zero--field case so that the monopole contribution and
$\chi$ are both zero.
 
We begin with (constant) harmonic potentials defined on the
$d$--dimensional torus $R^d/\Lambda$, where the lattice $\Lambda$ consists
of points of the form $\;L\sum n^j \vec E_j$, the $n^j$ being integers and
the $\;\vec E_j$ being $d\;$ linearly independent vectors in $R^d$.  It
is,
of course, possible to set $L=1$ but we prefer to keep this prefactor in
order to track the effect of scaling the lattice.  Note that a constant
field--free potential is not pure gauge since the ``would be'' gauge
transformation is not single valued.  These potentials are closed but not
exact.  Equivalently one sees that the Wilson loop $ W(\gamma_j) =
\exp(-i\oint_{\gamma_j} \vec A\cdot d\vec x) = \exp(-i\vec A\cdot
\vec E_j \; L)$
evaluated on the closed non--contractible loop $\gamma_j$ from $\vec 0$ to
$L\vec E_j$ is gauge invariant, and is therefore an obstruction to gauging
the potential away to zero.
 
The eigenvalues of $-D^2$ and $\dirac^2$ are in fact identical for zero
field. On the multidimensional torus with periodic boundary
conditions, $\psi(\vec x)=\psi(\vec x + \vec\mu)$,
$\vec\mu\in\Lambda$, the eigenvalues are easily computed to be
\begin{equation}
\lambda_n =  {\left({2\pi\over L}\right)^2} \cdot
  g^{ij}\,(n_i-a_i)(n_j-a_j);\qquad \vec n\in Z^d,
\label{E:eigen}
\end{equation}
where the matrix $(g^{ij})$ is the inverse of
$g_{ij}=\vec E_i\cdot\vec E_j$;  and $ a_i={L\over 2\pi}\vec E_i \cdot
\vec A$. To compute the corresponding effective action one defines
\begin{equation}
\Seff=\half\ln\det(-D^2)=-\half\left.{d\over ds} \z(s)\right\vert_{s=0,}
\label{E:Edetdef}
\end{equation}
where $\,\z(s)$ is the zeta function associated with $-D^2$,
\begin{equation}
\z(s)  = \tr'\left(-{D^2\over\mu^2}\right)^{-s} 
= {\sum}' \left({\lambda_n\over \mu^2}\right)^{-s}_.
\label{E:zeta}
\end{equation}
It should emphasised that the zeta function definition embodied in
(\ref{E:zeta}) implies both a regularization and a renormalization.
When comparing the zeta function result with effective actions
calculated using other renormalization prescriptions (\eg,
Schwinger's) one should always bear in mind that different
renormalization prescriptions will yield effective actions that may
differ by a finite renormalization. In particular, the quantity $\mu$
appearing in (\ref{E:zeta}) is a renormalization scale, which has been
introduced to keep the zeta function dimensionless, and the dependence
of $\Seff$ on $\mu$ corresponds to a finite renormalization. The
dependence on the renormalization scale is logarithmic and
proportional to the conformal anomaly (see \eg,
reference~\cite{bvwIII}). In order that dimensionful quantities may be
properly displayed throughout, we shall retain this normalization
scale in all our calculations.

In the present case the zeta function associated with the differential
operator $-D^2$ is given in terms of a relatively well understood special
function by
\begin{equation}
\z(s) =  2 \left({\mu L \over 2\pi} \right)^{2s} \z_E(s, \vec a).
\label{E:zeta-def}
\end{equation}
The prefactor 2 arises from summing over both particle and
antiparticle states, while the symbol $\z_E(s, \vec a)$ denotes the
generalized Epstein zeta function,~\cite{EpsteinI,EpsteinII} defined by
the sum
\begin{equation}
\z_E (s,\vec a)={\sum_{Z^d}}'\;
\left[g^{ij}(n_i-a_i)(n_j-a_j)\right]^{-s}_. 
\label{E:sum}
\end{equation}
(the prime indicates that one should sum over non--zero eigenvalues only).

The generalised Epstein zeta function is difficult to handle
explicitly, at least in higher dimensions. Fortunately, we can apply
the generalised Poisson resummation formula
\begin{eqnarray}
&&\Sum_{Z^d} \exp\left[-\pi g^{ij} (n_i-a_i)(n_j-a_j)\right] =
  \sqrt{\det g_{ij}} \cdot
  \Sum_{Z^d} \exp\left[-\pi g_{ij} m^i m^j - 2\pi i m^i a_i\right].
\nonumber\\
&&
\end{eqnarray}
Taking a Mellin transform of the above gives~\cite{EpsteinI,Weil}
\begin{equation}
\z_E(s,\vec a)=
{\Gamma ({d\over 2}-s)\over \Gamma (s)}\;\pi^{2s-{d\over2}}\;
\sqrt{\det[g_{ij}]}\;\;{\sum}'
\big(g_{ij}\,m^i m^j\big)^{s-{d\over 2}}\;e^{-2\pi i\,m^ia_i}.
\label{E:epstein}
\end{equation}
The zero--mode, $m_i =0$, is eliminated because the zeta function is
defined by analytic continuation in $s$. For large $s$ the zero--mode
makes no contribution. Equation (\ref{E:epstein}) is a generalisation
of the well--known functional equation (reflection formula) for
Riemann's (or rather Euler's) zeta function.~\cite{Euler} It is now
relatively simple to show that
\begin{eqnarray}
\label{E:zp}
\z_E(0,\vec a)&=&0
\\
\z_E'(0,\vec a)&=&\Gamma \big(d/2\big)\; \pi^{-{d\over 2}}\sqrt{\det[g_{ij}]}
\;\;{\sum_{Z^d}}'\left(g_{ij}\,n^i n^j\right)^{-{d\over 2}}\;
e^{\;-2\pi i\, n^ja_j}.
\nonumber
\end{eqnarray}
The special case $d=2$ has been extensively studied by
Kronecker,~\cite{Kronecker} and for that reason we shall refer to
$\z_E'(0,\vec a)$ as a generalised Kronecker sum.
 
Since $\z_E$ vanishes at the origin the prefactor $({2\pi / L})^2$ in
(\ref{E:zeta-def}) does not contribute to the effective action and
\begin{equation}
\Seff = \half \ln\det(-D^2)=-\z_E'(0,\vec a).
\label{E:effa}
\end{equation}
Note in particular that in this case the effective action is
independent of the renormalization scale $\mu$. This is actually a
rather deep result, related to the vanishing of the conformal anomaly
for zero electromagnetic field strength.~\cite{bvwIII} If we consider
$\dsq$ instead of $-D^2$, the only change is to multiply the above
with minus one-half the number of spinor components in $d$ dimensions.
This general result may be related to known results in one and two
dimensions.

In the simplest case, $d=1$, the sum in (\ref{E:zp}) is a familiar
trigonometric series.~\cite{Fourier} We find the effective action
\begin{equation}
\Seff = \half \ln\det(-D^2) = \ln\left(4\sin^2 {LA\over 2}\right).
\label{E:effao}
\end{equation}
Indeed, in one dimension one does not need the reflection formula.
The exact zeta function is just a sum of Hurwitz zeta functions
and the effective action can be computed directly. Using other
techniques,~\cite{Zuber} we may show that the effect of a mass
term is to change the determinant to 
\begin{equation}
\Seff = \half \ln\det(-D^2+m^2) 
=
\ln\left\{4\left[
\sin^2\left({LA\over2}\right) +\sinh^2\left({mL\over2}\right) 
\right]\right\}.
\end{equation}
 
Considering the next simplest case, we observe that in two dimensions
any lattice may be re--scaled to satisfy
\begin{equation} 
g=\left[
\begin{array}{cc}1&\Re(\tau)\\ \Re(\tau)&\tau\bar{\tau}\end{array}
\right],
\label{E:metric}
\end{equation}
where $\tau$ is a complex parameter known as the Teichm\"{u}ller
parameter.  The Kronecker sum becomes
\begin{equation} 
\z_E'(0,\vec a)={\Im(\tau)\over \pi}\;\; {\sum}'\;\;{e^{-2\pi i
(ma_1+na_2)}\over \vert m+\tau n\vert^2}.  
\label{E:Krontwo} 
\end{equation}
This sum can be expressed in terms of Riemann
theta--functions.~\cite{EpsteinI,KKK} We find the effective action
\begin{equation}
\z_E'(0,\vec a) = -2\;\log \left|{1\over \eta(\tau)}\;\;\;
\vartheta\left\lbrack
\,\begin{array}{c}\half+a_1\\ \half-a_2\end{array}
\right\rbrack (0|\tau)\right|.
\label{E:theta}
\end{equation}
Our theta--function conventions are those of Mumford.~\cite{Mumford}
This two--dimensional effective action is in fact well known in string
theory.~\cite{AlvarezGaume,AlvarezGaumeII}
 
Though the discussion has, for clarity, been given in terms of an
Abelian gauge potential, the extension to non--Abelian gauge
potentials is simple.  If the field strength is zero, we may use the
non--Abelian version of Stoke's theorem to deduce that the Wilson
loops $W(\gamma) = \tr(P\exp\oint_\gamma \vec A\cdot d\vec x)$ form a
representation of $H_1$, the first homology group.  This implies that
the $\vec A(x) = \lambda_a \vec A^a(x) $ may be gauge fixed to be
mutually commuting constant matrices.~\cite{bvw} The results of
this section then continue to hold provided one introduces an
additional product over the gauge group, $\prod_{a=1}^{dim(G)}$.

\section{Constant Field Strength: Two Dimensions}
\def\x{{\textstyle {1\over2} + {m^2\over2B} }}
\def\xx{{\textstyle 1 + {m^2\over2B} }}
\def\xxx{{\textstyle {m^2\over2B} }}
Consider the case of a constant electromagnetic field in two
dimensions.  Let the area of spacetime be denoted by $S$.  The field
strength may be written as:
\begin{equation}
F_{\mu\nu}=\left(\begin{array}{cc}0&B\\ -B&0\end{array}\right).
\label{E:strength2}
\end{equation}
As a potential we choose $ A_\mu =(0,B x) $ .
 
\no \underbar{Scalar Particles.}\Break
With this choice the gauged Laplacian appropriate to scalar particles
is
\begin{equation}
-D^2=-\curl^2_{x}-(\curl_{y}-iBx)^2.
\label{E:decomp}
\end{equation}
By observing that $-D^2$ commutes with the momentum $ {\hat P}_{y}=
-i\curl_{y} $ we see that on eigenstates of ${\hat P}_y$ it reduces to
\begin{equation}
-D^2 \to -\curl^2_{x}+B^2\left(x-{p\over B}\right)^2.
\end{equation}
Since this is just (twice) the Hamiltonian of a harmonic oscillator it
has eigenvalues $\lambda_{p,n}=(2n+1)|B|$.  These eigenvalues are
independent of $p$. Thus all levels are degenerate. Later we shall see
that this degeneracy is just $ 2(|B|\cdot S)/2\pi$, the prefactor 2
again arising from the sum over both particle and antiparticle states.
For notational simplicity take $B$ to be positive.
 
The zeta function of $-D^2+m^2$ is given in terms of a Hurwitz zeta
function by:
\begin{equation}
\z(s)= {2BS\over 2\pi} 
\Sum_{n=0}^\infty \left[{(2n+1)B+m^2\over\mu^2}\right]^{-s}
\equiv {2BS\over 2\pi} \left({2B\over\mu^2}\right)^{-s}\;\z_H(s;\x).
\label{E:3.4}
\end{equation}
 
\no The one--loop effective Lagrangian density is $\Leff = \Seff/S =
\half\ln\det(-D^2+m^2)/S = -\half\z'(0)/S$. Using properties of the
Hurwitz zeta function discussed in the appendix, in particular its
value and slope at the point $s=0$, yields:
\begin{equation}
\Leff = -{m^2\over4\pi}\cdot\ln\left({2B\over\mu^2}\right) -
      {B\over 2\pi}\cdot\ln\Big(\Gamma(\x)/\sqrt{2\pi}\Big).
\end{equation} 
It is easy to see that the $m\to0$ limit is well behaved --- $\Leff
\to (B/4\pi)\cdot\ln2$ --- while in the strong-field limit
\begin{equation}
\Leff = -{m^2\over4\pi}\cdot\ln\left({2B\over\mu^2}\right) 
 + {B\over 4\pi}\ln 2 + O(1).
\end{equation} 
This simple example is already a counterexample to the folk
theorem $\Leff\sim B^{d/2}\ln B$.

The weak-field $B\to0$ limit may be taken by making use of the
doubling formula for the Hurwitz zeta function:
\begin{equation}
\zeta_H\left(s;\half+{x\over2}\right) = 
2^s\; \zeta_H(s;x) - \zeta_H\left(s;{x\over2}\right),
\label{E:3.7}
\end{equation} 
which follows from the definition (\ref{E:A.1}) of the appendix. Using
this doubling formula we may write (\ref{E:3.4}) as
\bea \zeta(s) &=& {2BS\over2\pi} \left({2B\over\mu^2}\right)^{-s}
\zeta_H\left(\half+{m^2\over2B}\right)
\nonumber\\
&=& {2BS\over2\pi} \left[ \left({B\over\mu^2}\right)^{-s} \;
  \zeta_H\left(s;{m^2\over B}\right) -
  \left({2B\over\mu^2}\right)^{-s} \;
  \zeta_H\left(s;{m^2\over2B}\right) \right].  
\eea 
With the help of equations (\ref{E:A.9}), (\ref{E:A.11}), and
(\ref{E:A.18}), we may develop the weak-field expansion for the
effective Lagrangian:
\\
{\sf [there is a typo, an extaraneous 2, in the second line of the
  published version]}
\bea \Leff &=& - \half {\zeta'(0)\over S}
\nonumber\\
&=& - {B\over2\pi}\Bigg[ -\ln\left({B\over\mu^2}\right) \;
\zeta_H\left(0;{m^2\over B}\right) +\zeta'_H\left(0;{m^2\over
    B}\right)
\nonumber\\
&& + \ln\left({2B\over\mu^2}\right) \; \zeta_H\left(0;{m^2\over2
    B}\right) -\zeta'_H\left(0;{m^2\over2B}\right) \Bigg]
\nonumber\\
&=& - {B\over2\pi}\Bigg[ \ln\left({m^2\over\mu^2}\right) \;
\zeta_H\left(0;{m^2\over B}\right) + {m^2\over B} - {1\over12} {B\over
  m^2} \nonumber
\\
&& -\ln\left({m^2\over\mu^2}\right) \; \zeta_H\left(0;{m^2\over
    2B}\right) -{1\over2} {m^2\over B} + {1\over12} {2B\over m^2}
\nonumber
\\
&& + \sum_{k=1}^{n-1} { {\bf B_{2k+2}} \over (2k+2)(2k+1) }
\left({B\over m^2}\right)^{2k+1} \left\{ 2^{2k+1} - 1 \right\} +
O\left[\left({B\over m^2}\right)^n \right] \Bigg]
\nonumber\\
&=& {1\over2\pi} \Bigg[ {1\over2} m^2 \left\{ 1 -
  \ln\left({m^2\over\mu^2}\right)\right\} + {1\over12} {B^2\over m^2}
\nonumber\\
&& + m^2 \sum_{k=1}^{n-1} { {\bf B_{2k+2}} \over (2k+2)(2k+1) }
\left({B\over m^2}\right)^{2k+2} \left\{ 2^{2k+1} - 1 \right\} 
+ O\left[\left({B\over m^2}\right)^n \right] \Bigg]. \qquad 
\eea 
Here the symbol ${\bf B_{n}}$ denotes the $n$th Bernoulli number.
Notice that there are no terms logarithmic in $B$ in the weak-field
expansion.  However, as foreshadowed, the zeta-function
renormalization has introduced nonstandard finite terms into the
effective Lagrangian. An additional finite (often $\mu$-dependent)
renormalization is needed to remove these terms. Removing the constant
term in the above equation corresponds to renormalizing the effective
cosmological constant, while removing the term proportional to $B^2$
corresponds to a finite renormalization of electric charge. (We have
chosen our notation in such a manner that electric charge does not
appear explicitly.) Having implemented these additional finite
renormalizations we (finally) display the renormalized effective
action as an asymptotic series starting with $B^4$: 
\be \Leff =
{m^2\over2\pi} 
\sum_{k=1}^{n} { {\bf B_{2k+2}} \over (2k+2)(2k+1) }
\left({B\over m^2}\right)^{2k+2} \left\{ 2^{2k+1} - 1 \right\} 
+O\left( B^{2n+4} \right).  
\ee 
(We have gone through this simple example in admittedly tedious detail
because the same techniques will be used over and over in the following
discussion.)

In order to confirm the degeneracy factor we consider the
heat kernel of $-D^2+m^2$.  This heat kernel is found to be
\begin{eqnarray}
K(t)&\equiv& \tr'\left\{\exp([D^2-m^2]t/\mu^2\right\}
\nonumber\\
&\equiv& e^{-m^2t/\mu^2} \cdot
\Sum_{n=0}^\infinity
e^{-\lambda_n t/\mu^2}
\nonumber\\
&=&
{BS\over 2\pi}\cdot e^{-m^2t/\mu^2}\cdot
\cosech(Bt/\mu^2).
\label{E:above}
\end{eqnarray}
Note that as $t\to0$, $K(t)\to2(S\mu^2/4\pi t)$, as it should according
to the (known) short time behaviour of the heat kernel. This {\sl a
  posteriori} proves that we have chosen the correct degeneracy factor
$2(BS/2\pi)$. Equivalently, one may use the fact that as $B\to0$ the
heat kernel must approach that of the free two--dimensional diffusion
operator, to obtain an alternative verification of the degeneracy
factor. It should be emphasized that we are considering the heat
kernel appropriate to a \emph{complex} scalar field. That is to say,
the sum in equation (\ref{E:above}) includes both particle and
anti particle sectors.

\no\underbar{Dirac Particles.}\Break
The analysis for Dirac spinors closely parallels that of the charged
scalar field. The square of the Dirac operator is
\begin{equation}
\dsq =-D^2 + \Sigma_{\mu\nu}F^{\mu\nu}
     =-\curl^2_{x}-(\curl_{y}-iBx)^2 + \g5 B.
\end{equation}
The eigenvalues are $\lambda_n=(2n+1)|B| \pm B$. We now re--index the
states so that the eigenvalues become $\lambda_n = 2n|B|$, where the
effective degeneracy is $(|B|S/2\pi)$ for $n=0$ and $2(|B|S/2\pi)$ for
$n > 0$.
 
The zeta function of $\dsq$ is given in terms
of Hurwitz zeta functions (henceforth we take $B > 0$),
\begin{eqnarray}
\z(s) &=& 
{BS\over 2\pi} \left\{ 2\cdot\Sum_{n=1}^\infty 
\left({2nB+m^2\over\mu^2}\right)^{-s}
+\left({m^2\over\mu^2}\right)^{-s} \right\} 
\nonumber\\
&=& {BS\over 2\pi} \cdot \left\{ 2\cdot\left({2B\over\mu^2}\right)^{-s} 
\cdot \z_H(s;\xx)
+ \left({m^2\over\mu^2}\right)^{-s} \right\}
\nonumber\\
&=& {BS\over 2\pi} \cdot \left\{ 2\cdot\left({2B\over\mu^2}\right)^{-s} 
\cdot \z_H\left(s;{m^2\over2B}\right)
- \left({m^2\over\mu^2}\right)^{-s} \right\}
\end{eqnarray}
The one--loop effective Lagrangian density for spinors is $\Leff =
+\half\z'(0)/S$, so that:
\begin{equation}
\Leff= {B+m^2\over 4\pi}\ln\left({2B\over\mu^2}\right) 
+ {B\over 2\pi}\ln\Gamma(\xx)
      -{B\over 4\pi}\ln\left({2\pi\,m^2\over\mu^2}\right).
\label{E:3.14}
\end{equation} 
The weak--field limit $B\to 0$ for spinors is obtained just as it was
for scalars. As in the scalar case we must implement an additional
finite renormalization of the cosmological constant and the electric
charge. Then we obtain
\be 
\Leff =
{m^2\over4\pi} 
\sum_{k=1}^{n} { {\bf B_{2k+2}} \over (2k+2)(2k+1) }
\left({2B\over m^2}\right)^{2k+2}
+O\left(B^{2n+4} \right).  
\ee 
The limit $m\to0$ of equation (\ref{E:3.14}) is ill behaved ($\Leff
\to \infty$).  This is an infrared singularity associated with the
fact that the Dirac operator develops a zero--mode as $m\to0$. One may
deal with the zero--mode by simply dropping the ground state ($n=0$)
from the sum prior to taking the $m\to0$ limit. Equivalently, for
massless Dirac spinors one must exclude the zero mode ``by hand''.
Recall that the definition of the zeta function is in terms of $\sum'$
not $\sum$. The difference has up to now been irrelevant.  Taking
careful note of this difference leads to:
\begin{equation}
\z(s) = {2 BS\over 2\pi} \Sum_{n=1}^\infty \left({2nB\over\mu^2}\right)^{-s}
= {2BS\over 2\pi}  \left({2B\over\mu^2}\right)^{-s} \cdot \z_R(s).
\end{equation}
The one--loop effective Lagrangian density is simplified to
\begin{equation}
\Leff = +{B\over 2\pi}\cdot\ln\left({B\over\pi\mu^2}\right).
\end{equation}
 
The heat kernel of $\dsq+m^2$ is readily evaluated
\begin{equation}
K(t) = {BS\over 2\pi}\;
\coth\left({Bt\over\mu^2}\right)\;
\exp\left(-{m^2t\over\mu^2}\right).
\end{equation}
Note that as $t\to0$ that $K(t)\to2\cdot(S\mu^2/4\pi t)$, as indeed it
should.  (The factor of $2$ reflects the existence of two spinor
components in two dimensions).  In the massless case the
``zero--mode--suppressed'' heat kernel is
\begin{equation}
K'(t) = {BS\over 2\pi}\left[\coth\left({Bt\over\mu^2}\right)-1\right].
\end{equation}
Note that $K'(t)\to0$ as $t\to\infty$, thanks to the explicit exclusion
of the zero mode.
 
Finally, note that calculation of the heat kernels allows one
to deduce {\sl all} the Seeley--deWitt coefficients for the case of
constant field. Use of the Taylor series for $\cosech(Bt)$ and $\coth(Bt)$
yields:
\begin{eqnarray}
&a_{2n}(-D^2)& = -2 {(2^{2n-1} -1) \over (2n)!} \cdot {\bf B_{2n} }
  \cdot B^{2n}, \\
&a_{2n}(\dsq)& = {(2^{2n+1}) \over (2n)!} \cdot {\bf B_{2n} }
  \cdot B^{2n}, 
\end{eqnarray}
where ${\bf B_n}$ denotes the $n$th Bernoulli number. The `odd'
coefficients $a_{2n+1}$ all vanish. It is very unusual to know all the
$a_n$. For arbitrary fields these coefficients are leading order terms
in a low--momentum approximation. Thus they remain interesting for
general fields.

\section{Constant Field Strength: Three Dimensions}
In three dimensions the field strength may be written as:
\begin{equation}
F_{\mu\nu}=
\left(\begin{array}{ccc}0&0&0\\ 0&0&B\\ 0&-B&0\end{array}\right).
\label{E:strength3}
\end{equation}
Let the volume of spacetime be denoted by $V=S\cdot L$.
In the gauge $A_\mu = (0,0,By)$, we may immediately
write down the three--dimensional heat kernel $K_{d=3}(t) =
{\mu L\over\sqrt{4\pi t}}\cdot K_{d=2}(t)$.
Three--dimensional zeta functions quickly follow:
\begin{eqnarray}
\z_{d=3}(s) &=& {1\over\Gamma(s)}\cdot\int_0^\infty t^{s-1} K_{d=3}(t)
\nonumber\\
&=& {\mu L\over\sqrt{4\pi}}\cdot{\Gamma(s-\half)\over\Gamma(s)}\cdot
\z_{d=2}(s-\half)_.   
\end{eqnarray}
All factors are analytic at $s=0$. In particular, since
${1\over\Gamma(s)} \sim s + 0(s^2)$, the derivative at $s=0$ is
\begin{equation}
\z'_{d=3}(0) = {\mu L\over\sqrt{4\pi}}\cdot\Gamma(-\half)\cdot\z_{d=2}(-\half)
                = - \mu L \cdot\z_{d=2}(-\half).
\end{equation}
This means that the one--loop effective Lagrangian densities are given in
terms of $\z_H(-\half,x)$. Using results of the two--dimensional
discussion we see:
\begin{eqnarray}
\hbox{\rm Scalar Particles: \hfil}
&&\qquad \Leff = {1\over 4\pi}\cdot (2B)^{3/2} \cdot
   \z_H(-\half;\x), 
\nonumber\\
\hbox{\rm Dirac Particles: \hfil}
&&\qquad \Leff =  -{1\over8\pi}\cdot2B\cdot\left\{
   2 \sqrt{2B} \; \z_H(-\half;\xx) + m \right\}. 
\label{E:4.4}
\end{eqnarray}
Note that these effective actions are independent of the
renormalization scale $\mu$. This (nonobvious) result is a consequence
of the vanishing of the conformal anomaly in odd--dimensional
space--times.  For strong fields, the Hurwitz zeta function can be
computed by convergent series. [See equation (\ref{E:A.6}) of the
appendix.]  For the cases of interest:
\begin{eqnarray}
\z_H(-\half;\x) &=& \z_H(-\half;\half) - \Sum_{l=1}^\infty (-)^l
 {(2l-3)!!\over 2^l\;l!} \cdot
 \left({m^2\over2B}\right)^l \cdot \z_H(-\half+l;\half), 
\nonumber
\\
\z_H(-\half;\xx) &=& \z_R(-\half) -  \Sum_{l=1}^\infty   (-)^l
 {(2l-3)!! \over 2^l\;l!} \cdot
 \left({m^2\over2B}\right)^l \cdot \z_R(-\half+l). 
\label{E:4.5}
\end{eqnarray}
The coefficients in these expansions can be obtained (with severely
limited accuracy) from the tables of Jahnke and Emde.~\cite{Jahnke}
Alternatively, the proliferation of personal computers allows these
coefficients to easily be computed to any desired accuracy.
Note the presence of fractional powers of the field strength and the
absence of logarithmic terms when equation (\ref{E:4.5}) is
substituted into (\ref{E:4.4}) in order to obtain an expansion for the
effective Lagrangian.  We shall see that these general features
persist in any odd number of dimensions, at least for constant field
strength. The leading strong--field limit in three dimensions is
\begin{eqnarray}
\hbox{Scalar particles:} && \qquad
\Leff = {1\over4\pi} (2 B)^{3/2} \zeta_H\left(-\half;\half\right).
\nonumber
\\
\hbox{Spinor particles:} && \qquad
\Leff = - {1\over2\pi} (2 B)^{3/2} \zeta_R\left(-\half\right).
\label{E:4.6}
\end{eqnarray}
The leading term in the strong-field expansion for the spinor
Lagrangian has previously been obtained by Redlich.~\cite{Redlich1}

The leading coefficients appearing in the strong--field limit have
been evaluated numerically. This was done by using the reflection
formula for the Riemann zeta function to write $\z_H(-\half;1) =
\z_R(-\half)= -{1\over4\pi} \z_R(\threehalves)$. Note that
$\z_R(\threehalves)$ is given by a nicely convergent series suitable
for computer evaluation. We find $\z_R(\threehalves) \approx
2.612375$, $\z_R(-\half) \approx -0.207886$.  In a similar vein, we
use the ``doubling formula'', $\z_H(s;\half) = (2^s -1) \z_R(s)$ to
deduce
\begin{equation}
\z_H(-\half;\half) = {1\over4\pi} \left\{1 - {1\over\sqrt2}\right\}
\z_R(\threehalves) \approx 0.060888.
\end{equation}
 
In order to obtain the weak--field limit for the effective Lagrangian,
we return to equation (\ref{E:4.4}). We shall need equation
(\ref{E:A.7}) of the appendix, evaluated at $s=\half$.
\bea
\z_H\left(-{1\over2};x\right) &=& {1\over\Gamma(-\half)} \Bigg[
x^{3/2} \Gamma\left(-{3\over2}\right) 
+{1\over2} x^{1/2} \Gamma\left(-{1\over2}\right)
\nonumber\\
&& 
+\sum_{k=1}^n {\bf B_{2k}} {\Gamma(2k-{3\over2})\over(2k)!} x^{3/2-2k}
+ O(x^{1/2-2n}) \Bigg].
\eea

After an additional finite renormalization of the cosmological
constant and the electric charge, the weak-field limit for the scalar
effective Lagrangian is
\be
\Leff = -{m^3\over\pi} 
\sum_{k=1}^{n}  {\bf B_{2k+2}} 
{\Gamma(2k+\half)\over (2k+2)!\Gamma(-\half) }
\left[2^{2k}-{1\over2}\right] 
\left({B\over m^2}\right)^{2k+2}
+O\left(B^{2n+4} \right), 
\ee
while the spinor Lagrangian is
\be
\Leff = -{m^3\over4\pi} 
\sum_{k=1}^{n}  {\bf B_{2k+2}} 
{\Gamma(2k+\half)\over (2k+2)!\Gamma(-\half) }
\left({2B\over m^2}\right)^{2k+2}
+O\left(B^{2n+4} \right).  
\ee
Only even, positive-integer powers of the field enter the weak-field
expansion.

Notice the absence form our discussion of the Chern--Simons secondary
characteristic class (the topological mass term for the photon). This
term might be expected to appear on rather general grounds, following
arguments of Niemi and Semenoff~\cite{Niemi} and
Redlich~\cite{Redlich1}. However, a simple analysis is sufficient to
show that the Chern-Simons term vanishes for constant field
strength.~\cite{Redlich1,Redlich2} We should also note the appearance
of fractional powers of the field strength has also been noted in
Redlich's calculation for a massive spinor.~\cite{Redlich2} The
present calculation generalizes this result to the massive case and
also to scalar particles.

\section{Constant Field Strength: Four Dimensions}
In four dimensions the field strength may be written as:
\begin{equation}
F_{\mu\nu}=\left(\begin{array}{cccc}0&E&0&0\\ -E&0&0&0 \\
                         0&0&0&B\\  0&0&-B&0\end{array}\right),
\label{E:strength4}
\label{E:5.1}
\end{equation}
so that the four--dimensional problem breaks up into two
two--dimensional ones.

This result follows from the block-diagonalizability of antisymmetic
matrices through orthogonal transformations. It is important to know
that $E$ and $B$ are invariant scalars that characterise the
electromagnetic field.  Note that $E^2 + B^2 = \half F_{\mu\nu}
F^{\mu\nu} = \half F^2$, and $ 2EB = \eps_{\mu\nu\sigma\rho}
F^{\mu\nu} F^{\sigma\rho} = F\tilde F $.  While it is more common to
characterise the field in terms of $F^2$ and $F\tilde F$, it is more
useful for us to use the invariants $E$ and $B$.  Though we shall be
working in Euclidean space, we note that the result (\ref{E:5.1}) also
obtains in Minkowski space. This follows from the fact that it is
always possible to make a Lorentz boost such the electric and magnetic
fields become parallel and then (by a rotation) to make both fields
point in the $x$ direction.

\subsection{\underline{$E=0$, $B\neq0$}}
Our analysis in this case parallels that of three dimensions. Writing the
volume of spacetime as $\Omega=\Omega_\perp \cdot S$, and choosing the
gauge as $A_\mu=(0,0,0,By)$, we factorize the heat kernel
\begin{equation}
K_{d=4}(t) = D {\Omega_\perp \mu^2\over4\pi t} \cdot K_{d=2}(t).
\end{equation}
Here $D=1$ for scalars and $D=2$ for spinors, reflecting the fact that
four-dimensional spinors possess twice as many degrees of freedom as
two-dimensional spinors. The four--dimensional zeta function is given
by
\begin{equation}
\z_{d=4}(s) = {D\over\Gamma(s)}\cdot\int_0^\infty t^{s-1} K_{d=4}(t) dt =
 D {\Omega_\perp\mu^2\over4\pi} \cdot {\z_{d=2}(s-1)\over s-1}.
\end{equation}
Inserting the results of the two--dimensional case yields
\begin{eqnarray}
\hbox{\rm Scalar Particles:\hfill}
&&\qquad \z(s) = {B^2\Omega\over2\pi^2} \cdot 
\left({2B\over\mu^2}\right)^{-s} \cdot
          {\z_H(s-1;\x) \over (s-1)},   
\nonumber\\
\hbox{\rm Dirac Particles:\hfill}
&&\qquad \z(s) = {B^2\Omega\over\pi^2} \cdot 
\left({2B\over\mu^2}\right)^{-s} \cdot
          {\z_H(s-1;\xx) \over (s-1)}    
\nonumber\\
&&\qquad \qquad \qquad + {m^2 B\Omega\over 4\pi^2}
        \left\{ {(m/\mu)^{-2s} \over s-1} \right\}_.         
\end{eqnarray}
Taking derivatives at $s=0$ is straightforward.
 
\subsubsection{\underline{Scalar Particles:}}\Break
For the effective Lagrangian density one obtains:
\begin{equation}
\Leff = {B^2\over4\pi^2} 
\left\{ \left[1-\ln\left({2B\over\mu^2}\right)\right] \;\z_H(-1;\x) +
               \z_H'(-1;\x) \right\}.
\label{E:5.1.4}
\end{equation}
Using results from the appendix one finds that
\begin{eqnarray}
\Leff = {1\over4\pi^2} \Bigg\lbrack \left({B^2\over24} - {m^4\over8}\right) \cdot
 &&\left[1- \ln\left({2B\over\mu^2}\right)\right] 
+ B^2\z_H'(-1;\half) + {m^4\over8} 
\nonumber\\ +
&&  B^2 \int_0^{\xxx} \ln(\Gamma(\half+y)/\sqrt{2\pi}) dy. \Bigg\rbrack.
\end{eqnarray}
This form makes it easy to extract the large
field limit $B\gg m^2$,
\begin{equation}
\Leff = {1\over4\pi^2} \left\lbrack \left({B^2\over24} - {m^4\over8}\right) \cdot
  \left[1- \ln\left({2B\over\mu^2}\right)\right] 
+ B^2\z_H'(-1;\half) + {m^4\over8}  -
   {B m^2\over4} \ln2  \right\rbrack.\;
\end{equation}
The coefficient $\z_H'(-1;\half)$ has been evaluated numerically. This was
done by noting that $\z_H(s;\half) = (2^s -1) \cdot \z_R(s)$; and by using
the reflection formula (see appendix) to relate $\z_H'(-1;\half)$ to
$\z_R'(+2)$. We obtain  $\z_H'(-1;\half)\approx 0.053829$.

In order to obtain the weak-field limit it is easiest to return to
equation (\ref{E:5.1.4}) and to use the doubling formula (\ref{E:3.7})
along with equations (\ref{E:A.10}) and (\ref{E:A.16}) of the appendix.
This procedure yields an asymptotic expansion in terms of powers of
$B^2$; in particular there are no terms logarithmic in field strength.
After performing the usual additional finite renormalization of
cosmological constant and electric charge, we obtain the extended
Euler--Heisenberg effective Lagrangian
\bea
\Leff &=& {B^2\over16\pi^2} \Bigg[ {-7\over360} {B^2\over m^4} +
\sum_{k=1}^{n}  {{\bf B_{2k+2}} 
\over (2k+2) (2k+1)(2k) } \left[ 2^{2k+2} - 2 \right]
\left({B\over m^2}\right)^{2k}
+O\left(B^{2n+2}\right) \Bigg].  
\nonumber\\
&&
\eea
This result should be compared with the Minkowski-space
Euler--Heisenberg Lagrangian obtained by Schwinger.~\cite{SchwingerGauge}
In order to make the comparison, we must take Schwinger's formula and
after setting $E=0$ multiply by $-1$. In the general case, for which
$E$ does not vanish, we must replace the $E$'s in Schwinger's formula
with $iE$ before multiplying by $-1$ to obtain the Euclidean result.
Finally, we have implicitly chosen units for the renormalized charge
so that Schwinger's $\alpha^2$ corresponds to our $1/16\pi^2$. After
making these adjustments for notation and metric signature, we may
confirm that our expansion agrees with Schwinger to the leading order
(which is the only order explicitly displayed in
reference~\cite{Schwinger}. Note though, that in this zeta function
formalism, it is easy to display \emph{all} orders of the asymptotic
expansion for the Euler--Heisenberg Lagrangian in the weak-field
limit.

\par\penalty-500  
\subsubsection{\underline{Dirac Particles:}}\Break
For Dirac spinors one obtains
\begin{eqnarray}
\Leff &= -{B^2\over2\pi^2} 
\left\{ \left[1-\ln\left({2B\over\mu^2}\right)\right]\z_H(-1;\xx) +
               \z_H'(-1;\xx) \right\} 
\nonumber\\
&\qquad + {m^2B\over8\pi^2} 
\left[ \ln \left({m^2\over\mu^2}\right) -1 \right]. 
\label{E:5.1.8} 
\end{eqnarray}
A little work using results given in the appendix yields
\begin{eqnarray}
\Leff
&=& {1\over2\pi^2} \Bigg\{
\left({B^2\over12} + {m^2B\over4} +{m^4\over8}\right)
\left[1- \ln\left({2B\over\mu^2}\right)\right] - 
B^2\z_R'(-1) - {m^4\over8}  -{m^2B\over4}  
\nonumber\\
&&\qquad +{m^2B\over4}\left[\ln\left({m^2\over\mu^2}\right)-1\right] -
   B^2 \int_0^{\xxx} \ln(\Gamma(1+y)/\sqrt{2\pi}) dy \Bigg\}.
\end{eqnarray}
For strong fields one uses $\ln\Gamma(1+\eps) = -\gamma\eps
+O(\eps^2)$, to establish
\begin{eqnarray}
\Leff
&=& {1\over2\pi^2} \Bigg\{
\left({B^2\over12} + {m^2B\over4} +{m^4\over8}\right)
\left[1- \ln\left({2B\over\mu^2}\right)\right] 
- B^2\z_R'(-1) - {m^4\over8}  -{m^2B\over4}  
\nonumber\\
&&\qquad +{m^2B\over4}\left[\ln\left({m^2\over\mu^2}\right)-1\right] +
  {m^2B\over4}\ln(2\pi) +{\gamma\over8}m^4     \Bigg\}
   + m^4\; O(m^2/B). 
\end{eqnarray}
The coefficient $\z_R'(-1) \approx -0.165421$ has been calculated
numerically.
 
In order to obtain the weak-field limit we return to equation
(\ref{E:5.1.8}) and employ the identity $\zeta_H(s;1+x) = \zeta_H(s;x) -
x^{-s}$. Just as in the scalar case we obtain an (extended)
Euler--Heisenberg Lagrangian.
\bea
\Leff &=& {B^2\over16\pi^2} \Bigg[ {-2\over45} {B^2\over m^4} +
\sum_{k=1}^{n}  {4 \; {\bf B_{2k+2}} 
\over (2k+2) (2k+1)(2k) }
\left({2B\over m^2}\right)^{2k}
+O\left(B^{2n+2}\right) \Bigg],  
\nonumber\\
&&
\eea
which specializes to the Euclidean version of Schwinger's
result.~\cite{Schwinger}

The results of this section may easily be extended to discuss the case
$B=0$, $E\neq0$. One needs merely replace $B$ by $E$ in all formulae
to obtain valid Euclidean space results. Recall, though, that in
continuing to Minkowski space one effects the transformation $E\to i
E$. In the next section we shall begin to address the complexities
encountered when $E$ and $B$ are both nonzero.
 
\subsection{\underline{$E\neq 0, B\neq 0$}}
The scalar spectrum is given by $\lambda = (2n+1)|E|+(2n'+1)|B|+m^2$,
where all the states have the degeneracy $2(|EB|\Omega)/4\pi^2$. The
zeta function is
\begin{equation}
\z(s) = {|EB|\Omega\over2\pi^2}\cdot \Sum_{n,n'=0}^{\infty}
         \left[{ (2n+1)|E|+(2n'+1)|B|+m^2 \over \mu^2}\right]^{-s}.
\label{E:Escalar}
\end{equation}
For spinors the analogous zeta function is a trifle more complicated
[{\tt the published version of this equation has a typo}]:
\begin{eqnarray}
\z(s) = {|EB|\Omega\over4\pi^2}\cdot\Bigg\{&&
    4\cdot\Sum_{n,n'=1}^{\infty}
         \left[ {2n|E|+2n'|B|+m^2 \over\mu^2} \right]^{-s} 
\nonumber\\
    &&+2\cdot\Sum_{n=1}^\infty \left[{2n|E|+m^2\over\mu^2}\right]^{-s} 
\nonumber\\
    &&+2\cdot\Sum_{n'=1}^\infty \left[{2n'|B|+m^2\over\mu^2}\right]^{-s} 
\nonumber\\
    &&+1\cdot \left[{m^2\over\mu^2}\right]^{-s}  \Bigg\}.
\label{E:Espinor}
\end{eqnarray}
These zeta functions are in general so unwieldy as to be unmanageable.
To proceed we restrict ourselves to the case $|E|=|B|=F$,
corresponding to self--dual and anti--self--dual constant fields.  In
this special situation we can recast these zeta functions in terms of
the elementary Hurwitz zeta functions.
 
\def\X{{ \textstyle {1\over2} + {m^2\over2F} }}
\def\XX{{ \textstyle 1 + {m^2\over2F} }}
\def\Xb{{ \textstyle {m^2\over2F} }}
For the case of a scalar particle
\begin{equation}
\z(s) = {F^2\Omega\over2\pi^2}\cdot 
\left({2F\over\mu^2}\right)^{-s}\cdot \Sum_{n,n'=0}^\infty
         \left( n+n'+1+{\t{m^2\over2F}} \right)^{-s}
\end{equation}
We define $t=n+n'$, and observe that there are precisely $t+1$ ways
in which $n$ and $n'$ can be arranged to sum to $t$, consequently
\begin{eqnarray}
\z(s)& =& 
{F^2\Omega\over2\pi^2}\cdot \left({2F\over\mu^2}\right)^{-s}\cdot 
\Sum_{t=0}^{\infty}
         (t+1)\cdot \left( t+1+{\t{m^2\over2F}} \right)^{-s} 
\\
      & = &{F^2\Omega\over2\pi^2}\cdot \left({2F\over\mu^2}\right)^{-s}
\cdot \left\{
         \z_H(s-1;\XX) -{\t {m^2\over2F}}\cdot \z_H(s;\XX) \right\}
\\
&=& 
{F^2\Omega\over2\pi^2}\cdot \left({2F\over\mu^2}\right)^{-s}
\cdot \left\{
         \z_H(s-1;\Xb) -{\t {m^2\over2F}}\cdot \z_H(s;\Xb) \right\}.
\end{eqnarray}
Here we have used $\zeta_H(s;1+x) = \zeta_H(s;x) - x^{-s}$.  For the
Dirac spinor, very similar results obtain, though one must be careful
with degeneracy factors:
\begin{eqnarray}
\z(s) = {F^2\Omega\over4\pi^2} \Bigg\{
&&\left({2F\over\mu^2}\right)^{-s}\cdot4\Sum_{n,n'=1}^\infty (n+n'+{\t{m^2\over2F}})^{-s} 
\\
+&&\left({2F\over\mu^2}\right)^{-s}\cdot2\cdot2\Sum_{n=1}^\infty (n+{\t{m^2\over2F}})^{-s}
+\left({m^2\over\mu^2}\right)^{-s} \Bigg\}_. 
\end{eqnarray}
Re--indexing the sum, using $t=n+n'-1$, yields
\begin{equation}
\z_{\rm Spinor}(s) = 
2 \; \z_{\rm Scalar}(s) 
+ {F^2\Omega\over4\pi^2}\left({m^2\over\mu^2}\right)^{-s}.
\end{equation}
Apart from the ground--state contribution ($n=n'=0$; $\lambda =
m^2$) the scalar and spinor zeta functions are proportional to one
another. This desirable property does not, unfortunately, continue to hold
if $|E|\neq|B|$. To discuss the effective action, the strong--field limit,
and the weak--field limit, it suffices to discuss the scalars.
 
The effective Lagrangian for scalars is
\begin{eqnarray}
\Leff = {F^2\over4\pi^2} \Bigg\{ \ln\left({2F\over\mu^2}\right)
&&\left[\z_H(-1;\Xb) - {\t{m^2\over2F}} \z_H(0;\Xb) \right] \\
&&-\z'_H(-1;\Xb) + {\t{m^2\over2F}}\z'_H(0;\Xb) \Bigg\}.
\end{eqnarray}
Using results from the appendix yields
\begin{eqnarray}
\Leff &= & -{1\over4\pi^2}
     \left[{1\over12} F^2 -{{1\over8}}m^4\right]\ln\left({2F\over\mu^2}\right) \\
&& -{F^2\over4\pi^2} \cdot \z_R'(-1) \\
&& -{1\over4\pi^2}\left\{\quarter m^2 F + {\t{1\over8}} m^4 -
  \half m^2 F  \ln\left({\Gamma(\XX)\over\sqrt{2\pi}}\right) \right\} \\
&& -{F^2\over4\pi^2}\cdot
   \int_0^{m^2\over2F} \ln(\Gamma(1+y)/\sqrt{2\pi}) dy.
\end{eqnarray}
The strong--field and weak--field limits may be written down in the
standard manner, we omit details and present the results. For strong
fields
\begin{eqnarray}
\Leff = -{1\over4\pi^2}
\Bigg\{&&\left[\twelfth F^2 -{\t{1\over8}}m^4\right] 
\ln\left({2F\over\mu^2}\right)
      +F^2 \cdot \z_R'(-1) 
\\
&& +\quarter m^2 F + {{1\over8}}(1+\gamma)m^4 + 
O\left({m^6\over F}\right ) \Bigg\}.
\end{eqnarray}
For weak fields, the Euclidean-space (scalar) Euler--Heisenberg
Lagrangian is
\bea
\Leff &=& {F^2\over16\pi^2} \Bigg[ {-1\over15} {F^2\over m^4} +
\sum_{k=1}^{n}  {{\bf B_{2k+2}} 
\over (2k+2) (2k+1) } \left\{4+{2\over k} \right\} 
\left({2F\over m^2}\right)^{2k}
+O\left(F^{2n+2}\right) \Bigg].
\nonumber\\
&&
\eea
When continued to Minkowski space, the lowest term in this expansion
reproduces Schwinger's result.~\cite{Schwinger} To relate this scalar
result to the case of a massive Dirac spinor, one notes that
\begin{equation}
{\script L}_{\rm Spinor} = -2\cdot {\script L}_{\rm Scalar}
   + {F^2\over8\pi^2} \cdot \ln(m^2/\mu^2).
\end{equation}

The massless case affords considerable simplifications; for the scalar one
obtains
\begin{equation}
\z(s) = {F^2\Omega\over2\pi^2}\cdot\left({2F\over\mu^2}\right)^{-s}\cdot 
\z_R(s-1).
\end{equation}
This leads to an effective action
\begin{equation}
\Leff = -{F^2\over4\pi^2} \cdot
          \left[\twelfth\ln\left({2F\over\mu^2}\right) + \z_R'(-1) \right].
\end{equation}
The term proportional to $F^2$ is an artifact of the zeta-function
method, and may, in the usual fashion, be removed through a finite
renormalization of the electric charge.  For massless Dirac spinors,
explicit exclusion of the zero--mode leads to ${\script L}_{\rm
  Spinor} = -2{\script L}_{\rm Scalar}$. These massless
(anti--)self--dual effective actions have been previously discussed in
the literature, see for example~\cite{Notes,Elizalde,Soto}.

\section{Even Dimensionality}
In $d=2N$ dimensions a constant field may be brought into the block
diagonal form
\begin{equation}
F_{\mu\nu}=\left(
\begin{array}{ccccccc}0&B_1&&&&&\\ -B_1&0&&&&&\\&&\cdot&&&&\\
&&&\cdot&&&\\ &&&&\cdot&&\\
&&&&&0&B_N\\&&&&&-B_N&0\end{array}
\right),
\label{E:strength-E}
\end{equation}
where the $\pm B_j$ are the zeros of the characteristic polynomial
$\det(B\cdot I +iF)$. The problem thus decomposes into $N$ two--dimensional
problems, allowing the associated zeta functions to be written down by
inspection.
 
\no \underbar{Scalar Particles}\Break
The eigenspectrum is $\lambda_{\vec n} = \{\sum_{i=1}^N
(2n_i+1)|B_i|\} + m^2$ with degeneracy $d_n = \prod_{i=1}^N 2
(|B_i|\cdot S_i/2\pi)$. The zeta function is
\begin{eqnarray}
\z(s) &=&   2 \prod_{i=1}^N \left({|B_i|S_i\over2\pi}\right)
     \cdot \Sum_{n_i=0}^\infty \left({\{\sum_{i=1}^N (2n_i+1)|B_i|\} +
                     m^2\over\mu^2}\right)^{-s} \\
               &=& 2{\Omega\over\Gamma(s)}\cdot\int_0^\infty
                        t^{s-1}
  \left\{\prod_{i=1}^N \left({|B_i|\over4\pi}\right) 
\; \cosech\left({|B_i| t\over\mu^2}\right) \right\}
                      \exp(-{m^2 t\over\mu^2}).
\end{eqnarray}
 
\no \underbar{Dirac Particles}\Break
The eigenspectrum is $\lambda_{\vec n} = \{\sum_{i=1}^N 2n_i|B_i|\} +
m^2$ with degeneracy $d_n = \prod_{i=1}^N (|B_i|\cdot S_i/2\pi)\cdot
2^{\omega(n_i)}$,where $\omega(0)=0$, and $\omega(n>0) =1$.
The zeta function is
\begin{eqnarray}
\z(s) &=&   \prod_{i=1}^N \left({|B_i|S_i\over2\pi}\right)\cdot
      \Sum_{n_i=0}^\infty 2^{\omega(\vec n)}
        \left({\{\sum_{i=1}^N
                2n_i|B_i|\} +
                     m^2\over\mu^2} \right)^{-s} \\
               &=& {\Omega\over\Gamma(s)}\cdot\int_0^\infty
                        t^{s-1}
 \left\{\prod_{i=1}^N \left({|B_i|\over2\pi}\right) 
\coth\left({|B_i| t\over\mu^2}\right)  \right\}
                      \exp\left(-{m^2 t\over\mu^2}\right)_.
\end{eqnarray}
 
These zeta functions are tedious to calculate with in general, to
proceed we make the radically simplifying assumptions that
$|B_1|=|B_2|\!=\ldots=\!B$, and that $m=0$. Of course, in two
dimensions the condition on the field is always met since there is
only one $B$. In four dimensions it means that we confine ourselves to
(anti--)self--dual backgrounds.  With these assumptions the scalar
zeta function is simplified to
\begin{equation}
 \z(s) = \left({BS\over4\pi}\right)^N \cdot 
{(B/\mu^2)^{-s}\over\Gamma(s)} \cdot
            \int_0^\infty t^{s-1} (\cosech\, t)^N
\label{E:scalarZeta}
\end{equation}
while for the massless spinor, exclusion of the zero mode leads to
\begin{equation}
\z(s) = \left({BS\over2\pi}\right)^N \cdot {(B/\mu^2)^{-s}\over\Gamma(s)} 
\cdot
            \int_0^\infty t^{s-1} \{(\coth\, t)^N-1\}.
\label{E:spinorZeta}
\end{equation}
Note that for $N=2$, (\ie, four dimensions),  the spinorial zeta function is
four times the scalar zeta function, in agreement with our earlier
result.

For these backgrounds we can relate the zeta function in
$d$--dimensions to the ones in $(d-2)$ and $(d-4)$ dimensions by
recursion relations. For the scalar case, a repeated integration by
parts establishes
\begin{equation}
\z_N(s)=\left({\mu^2 S \over 4\pi}\right)^2 \cdot {\z_{N-2}(s-2)\over(N-1)(N-2)}
-\left({BS\over4\pi}\right)^2 \cdot
\left({N-2\over N-1}\right)\cdot\z_{N-2}(s),
\label{E:recursionI}
\end{equation}
so that for the scalar there are two disconnected series starting with $d=2$,
$d=4$, respectively. For the spinorial case a single integration by parts
suffices to obtain
\begin{equation}
\z_N(s)={\mu^2 S \over 2\pi} \cdot {\z_{N-1}(s-1)\over N-1}
+\left({BS\over2\pi}\right)^2\cdot\z_{N-2}(s).
\label{E:recursionII}
\end{equation}
In this case there is a single recursive series. We define
$\z_{N=0}(s) = 0$, while our previous calculations have shown
$\z_{N=1}(s) = {BS\over\pi}\cdot (2B/\mu^2)^{-s}\cdot \z_R(s)$. So we
deduce $\z_{N=2}(s) = ({BS\over\pi})^2 \cdot (2B/\mu^2)^{-s} \cdot
\z_R(s-1)$, which verifies our previous four--dimensional calculation.
 
Without resorting to numerical computation, we can make some general
observations regarding the $d$--dimensional case. Note that all the
$d$--dimensional zeta functions are of the form
\begin{equation}
\z_{d=2N}(s) = B^{d/2} \cdot \left({B\over\mu^2}\right)^{-s} \cdot\Omega\cdot  f(s)
\label{E:genone}
\end{equation}
in which case
\begin{equation}
\Leff \propto B^{d/2} \{ \ln (B/\mu^2) \cdot f(0) - f'(0)\}
\label{E:gentwo}
\end{equation}
so that the ``folk theorem'' generically holds for this class of
fields.  The ``folk theorem'' can fail if $f(0)$ happens to be zero;
this is in fact exactly what happens for massless scalar particles in
two dimensions. We remind the reader that the integrals
(\ref{E:scalarZeta}) and (\ref{E:spinorZeta}) defining $f(s)$ make sense
only for $\Re(s)>d/2$. The value of $f(0)$ is defined by analytic
continuation in $s$.
 
We also wish to point out that our recursive formulas
(\ref{E:recursionI}) and (\ref{E:recursionII}) are not the only way of
proceeding. The spectra are in this case sufficiently simple that
explicit calculation in terms of Hurwitz zeta functions is possible.
 
\section{Odd Dimensionality}
 
In $d=2N+1$ dimensions, we write the volume of spacetime as $\Omega_d
= L \cdot \Omega_{2N}$. The heat kernel is related to that in $2N$
dimensions by $K_d(t) = {L\over\sqrt{4\pi t}} \cdot K_{2N}(t)$, so
that (paralleling the discussion of the three--dimensional case) one
deduces
\begin{equation}
\z_d(s) = {L\over\sqrt{4\pi}}\cdot
{\Gamma(s-\half)\over\Gamma(s)} \cdot \z_{2N}(s-\half).
\end{equation}
Consequently
$\z'_d(0) = -L\cdot \z_{2N}(-\half)$, and in the notation of
equations (\ref{E:genone}) and (\ref{E:gentwo}) one has
\begin{equation}
\Leff \propto B^{d/2} \cdot f(-\half).
\end{equation}
Though this calculation is carried out for the special case $|B_1|=
\cdots = |B_i| = \cdots = |B_N| = B$, we expect that for general
fields the result $\Leff \sim B^{d/2}$ will remain true in any odd
number of dimensions.

\section{Conclusions}
 
We have studied the effect of constant field strengths on the
determinants of the Dirac operator and gauged Laplacian.  We began
with the topologically interesting cases of harmonic gauge potentials,
before proceeding to monopole potentials with constant field
strengths.  In both these cases we obtained the spectrum of the
squared Dirac operator and gauged Laplacian explicitly and calculated
the corresponding zeta function directly.  Computation of the
effective action then reduces to a relatively simple application of
special functions.
 
\appendix
\section{Hurwitz zeta Functions.}
 
To make this paper as self contained as possible, we shall present
here a number of definitions and useful results related to
zeta--function theory.  Most of these results may be obtained by
consulting Gradshteyn and Ryzhik.~\cite{Ryzhik} Other useful references
include Weil~\cite{Weil}, Abramowitz and
Stegun,~\cite{AbramowitzStegun} The Encyclopedia of
Mathematics,~\cite{Encyclopedia} and the Bateman Manuscript
Project.~\cite{Bateman}
 
The Hurwitz zeta function is defined by:
\begin{equation}
\z_H(s;x) = \Sum_{n=0}^\infty (n+x)^{-s}.
\label{E:A.1}
\end{equation}
This series converges absolutely for $\Re(s) > 1$, and the function so
defined may be analytically continued to the entire complex plane.
There is a single simple pole at $s=1$ and in that neighborhood
$\z_H(s,x) = {1\over s-1} -\psi(x) + o(s-1)$. Setting $x=1$ reproduces
the ordinary Riemann zeta function,
\begin{equation}
\z_R(s) = \z_H(s;1) = \Sum_{n=1}^\infty n^{-s}.
\label{E:A.2}
\end{equation}
The Riemann eta function is closely related to the Riemann zeta function.
It is defined by
\begin{equation}
\eta(s) = \Sum_{n=1}^\infty (-1)^{n+1} n^{-s}.
\label{E:A.3}
\end{equation}
The Riemann eta function, being defined by an alternating series, is
numerically much better behaved than the Riemann zeta function, and
furthermore converges over a larger region of the complex $s$--plane
[$\Re(s)> 0$]. In terms of the eta function
\begin{equation}
\z_R(s) = {1\over1-2^{1-s}}\cdot\eta(s).
\label{E:A.4}
\end{equation}
The Riemann zeta function satisfies the classical reflection formula
\begin{equation}
\z_R(s) = {(2\pi)^s\over\pi} \cdot \sin({\pi s\over2}) \cdot \Gamma(1-s)
          \cdot \z_R(1-s).
\label{E:A.5}
\end{equation}
By combining the reflection formula with an improved series representation
for the zeta function given in terms of the eta function one may {\sl
numerically} compute Riemann's zeta function over the entire complex plane.
 
The Hurwitz zeta function may be evaluated in terms of Riemann's zeta
function by using the binomial series to write
\begin{eqnarray}
\z_H(s;1+x) &=& \Sum_{n=1}^\infty \Sum_{l=0}^\infty
\left( {-s\atop l} \right) \cdot n^{-s-l} \cdot x^l 
\nonumber\\
&=& \z_R(s) + \Sum_{l=1}^\infty (-)^l
\cdot {s(s+1)\cdots(s+l-1)\over l!}
            \cdot      \z_R(s+l) \cdot x^l.  \qquad 
\label{E:A.6}
\end{eqnarray}
This series is convergent for $|x|<1$.  An asymptotic expansion for
large $x$ is provided by reference~\cite{Bateman} [equation 1.18 (9)]:
\begin{eqnarray}
\z_H(s;1+x)
&=& {1\over\Gamma(s)}\Bigg[ x^{1-s} \Gamma(s-1) + \half x^{-s} \Gamma(s) 
\nonumber\\
&&+ \Sum_{k=1}^n {\bf B_{2k}} {\Gamma(s+2k+1)\over (2k)!} x^{1-2-2k}
+ O(x^{-1-s-2n}) \Bigg].
\label{E:A.7}
\end{eqnarray}

\def\dummy{
*** surplus ***
For $|x|\ge1$ one uses the fact
that $\z_H(s;1+x) = \z_H(s;1+\frac(x)) -
\Sum_{n=1}^{\mathrm{integer}(x)} (n+\frac(x))^{-s}$. This result also
allows us to estimate the Hurwitz zeta function for large $x$ as
\begin{eqnarray}
\z_H(s;1+x)
&=& -\sum_{n=1}^{\mathrm{integer}(x)} (n+\frac(x))^{-s} + o(1) 
\\
&=& -\int_1^x y^{-s} dy +o(x^{-s}) = -{x^{1-s}-1 \over 1-s}
+O(x^{-s}).\\
\end{eqnarray}

*** surplus ***
}

At special values of $s$ more information is available.  The value of
the Hurwitz zeta function at non--positive integers is known in terms
of the Bernoulli polynomials
\begin{equation}
\z_H(-n;x) = -{{\bf B_{n+1}}(x)\over n+1} = 
-{{\bf B'_{n+2}}(x)\over(n+1)(n+2)},
\label{E:A.8}
\end{equation}
in particular,
\begin{equation}
\z_H(0;x) = \half-x;
\label{E:A.9}
\end{equation}
\begin{equation}
\z_H(-1;x) = -\half x^2 +\half x - \twelfth =
     -\half(x-\half)^2 + \twentyfourth.
\label{E:A.10}
\end{equation}
The derivative of the Hurwitz zeta function at $s=0$ is known to be
\begin{equation}
\z'_H(0;x) = \ln(\Gamma(x)/\sqrt{2\pi}).
\label{E:A.11}
\end{equation}
Derivatives at other values of $s$ are not given in the standard tables. One
may make some progress by noting that
\begin{equation}
{\curl\z_H(s;1+x)\over\curl x}  = -s \cdot \z_H(s+1;1+x)
\label{E:A.12}
\end{equation}
so that
\begin{equation}
{\curl\z'_H(s;1+x)\over\curl x} = -\z_H(s+1;1+x) - s\cdot\z'_H(s+1;1+x)_.
\label{E:A.13}
\end{equation}
This recursion relation, when applied to the case $s=-1$ yields
\begin{eqnarray}
\z'_H(-1;1+x)
&=& \z'_R(-1) + \half(x^2+x) +
      \int_0^x \ln(\Gamma(1+y)/\sqrt{2\pi}) dy 
\nonumber\\
&=& \z'_R(-1) + \half(x^2+x) -\half x\ln(2\pi) + x\ln\Gamma(1+x) 
\nonumber\\
&&\qquad\qquad\qquad - \int_0^x y \; \psi(1+y) \; dy.
\label{E:A.14}
\end{eqnarray}
This is the best analytic result that we have been able to obtain. For
$|x|<1$ we may make the convergent expansion
\begin{eqnarray}
\z'_H(-1;1+x) &=&\z'_R(-1) +\half(x^2+x) -\half x \ln(2\pi) +
                   x\ln\Gamma(1+x) 
\nonumber\\
              &&\quad+\half\gamma x^2 - \Sum_{k=2}^\infty (-)^k
                {\z_R(k)\;x^{k+1}\over (k+1)}. 
\label{E:A.15}
\end{eqnarray}
For $|x|$ large, a computationally useful asymptotic expansion may be
found in expositions by Elizalde and Soto~\cite{Elizalde,Soto}:
\begin{eqnarray}
\z'_H(-1;x) &=& (\half x^2 -\half x + \twelfth ) \cdot \ln x -\quarter x^2
 +\twelfth 
\nonumber\\
&& \qquad 
- \Sum_{k=1}^{n-1} {{\bf B_{2k+2}}\over (2k+2)(2k+1)(2k) }
                       \cdot x^{-2k} + O(x^{-2n}).
\label{E:A.16}
\end{eqnarray}
(But note theat the final sign is erroneously displayed in
reference~\cite{Elizalde}.)  We have resorted to numerical methods to
evaluate
\begin{eqnarray}
\z_R({\t {3\over2}}) \approx 2.612375;\qquad
  &&\z_R(-\half) \approx -0.207886; 
\nonumber\\
\z'_R(2) \approx -0.937548; \qquad
  &&\z'_R(-1) \approx -0.165421. 
\label{E:A.17}
\end{eqnarray}
 
Finally, we note that $\Gamma(\half) = \sqrt\pi$, and that Stirling's
approximation is
\begin{equation}
\ln(\Gamma(x)/\sqrt{2\pi}) = (x-\half)\ln(x) - x + {1\over12x} 
+ \Sum_{k=1}^{n-1} {{\bf B_{2k+2}} \over (2k+2)(2k+1)} \; x^{-1-2k} 
+O(x^{-2n}). \qquad
\label{E:A.18}
\end{equation}
It is often sufficient to use the simpler form
\begin{equation}
\ln(\Gamma(x)/\sqrt{2\pi}) = (x-\half)\ln(x-1) - (x-1) + O(1/x).
\label{E:A.19}
\end{equation}


\end{document}